\documentstyle[11pt,aaspp4,flushrt]{article} \parskip 0pt

\let\deg=\arcdeg
\let\mic=\micron
\def\Ref{\reference{}}

\def\about    {\hbox{$\sim$}}
\def\Lo       {\hbox{$L_\odot$}}
\def\Mo       {\hbox{$M_\odot$}}
\def\Mdot     {\hbox{$\dot M$}}
\def\k        {\hbox{$\kappa_\lambda$}}
\def\q        {\hbox{$q_\lambda$}}
\def\al       {\hbox{$\varpi_\lambda$}}
\def\lFl      {\hbox{$\lambda F_\lambda$}}
\def\tV       {\hbox{$\tau_V$}}
\def\tVo      {\hbox{$\tau_{V0}$}}
\def\Tsub     {\hbox{$T_{\rm sub}$}}
\def\am       {\hbox{$a_-$}}
\def\ap       {\hbox{$a_+$}}
\def\rgd      {\hbox{$r_{gd}$}}
\def\EBV      {\hbox{$E(B - V)$}}
\def\EV       {\hbox{$E(V - 12)$}}

\begin{document}

\rightline{To appear in \em ApJ Letters}

\title
                {ON PROTOSTELLAR DISKS IN HERBIG Ae/Be STARS}

\author{Anatoly Miroshnichenko\altaffilmark{1}, \v{Z}eljko
   Ivezi\'{c}\altaffilmark{2} and Moshe Elitzur\altaffilmark{3} \\
   Department of Physics and Astronomy, University of Kentucky, Lexington,
   KY 40506-0055}

\altaffiltext{1}{Permanent address: Pulkovo Observatory, St. Petersburg
                 196140, Russia; e-mail: anat@pulkovo.spb.su}
\altaffiltext{2}{e-mail: ivezic@pa.uky.edu}
\altaffiltext{3}{e-mail: moshe@pa.uky.edu}

\begin                           {abstract}

The spectral shape of IR emission from Herbig Ae/Be stars has been invoked as
evidence for accretion disks around high-mass protostars.  Instead, we present
here models based on spherical envelopes with $r^{-1.5}$ dust density profile
that successfully explain the observed spectral shapes.  The spectral energy
distributions (SEDs) of eight primary candidates for protostellar disks are
fitted in detail for all wavelengths available, from visual to far IR.  The
only envelope property adjusted in individual sources is the overall visual
optical depth, and it ranges from 0.3 to 3.  In each case, our models properly
reproduce the data for both IR excess, visual extinction and reddening. The
success of our models shows that accretion disks cannot make a significant
contribution to the radiation observed in these pre-main sequence stars.

\end{abstract}


\section
                                {INTRODUCTION}

Disks are expected to form during protostellar collapse, and low-mass stars
seem to provide good observational evidence for the existence of disks --- a
large body of T Tauri observations is successfully integrated in a model based
on circumstellar disks (Bertout \& Basri 1991).  On the other hand, in
high-mass stars, which form under different conditions with a strong tendency
to cluster, the observational evidence for disks is less compelling.  Based on
the SEDs of a sample of 47 Herbig Ae/Be stars, the high-mass ($\ga$ 1.5 \Mo)
counterparts of T Tauri stars, Hillenbrand et al.\ (1992, HSVK hereafter)
proposed that most of these objects too are surrounded by massive accretion
disks.  The SEDs were classified into three groups and HSVK suggested that the
IR emission from the 30 objects in group I is generated purely in optically
thick accretion disks. However, this proposal encountered some serious
difficulties, summarized by Evans \& Di Francesco (1995).  In particular, from
far-IR imaging of group I sources Di Francesco et al.\ (1994) conclude that
optically thick disks cannot account for the observed far-IR emission and that
another component is required, most likely a circumstellar envelope.

Hillenbrand et al.\ invoked the IR spectral behavior $\lFl \propto
\lambda^{-4/3}$ as the primary signature of optically thick disks.  However,
the emission from a spherical envelope with the density law $\rho(r) \propto
r^{-p}$ obeys $\lFl \propto \lambda^{-(p - 1)(\beta + 4)/2}$ if the envelope
is optically thin in the wavelength regime where the opacity varies according
to $\k \propto \lambda^{-\beta}$ (Harvey et al.\ 1991).  Interstellar grains
have $\beta$ = 1--2 at IR wavelengths, thus the $\lFl \propto \lambda^{-4/3}$
behavior is reproduced with $p$ \about\ 1.5, the expected spectral index of
the density profile during the spherical collapse phase (e.g.\ Shu, Adams \&
Lizano 1987).  Indeed, Hartmann, Kenyon \& Calvet (1993, HKC hereafter)
produced models for spherical envelopes with $r^{-1.5}$ density profile that
closely resemble the observed SEDs of Ae/Be stars longward of \about\ 3~\mic.
But the flux emerging in the visible region was too small in these models and
HKC suggested that there is a clear line of sight to the star through the
envelope, perhaps cleared by bipolar outflows.  The large axial cavities
required by this proposal somewhat blur the distinction between disk and
spherical geometries.

A prerequisite for settling this important controversy is a definitive answer
to the question: are spherical distributions at all capable of producing the
observed SEDs of Herbig Ae/Be stars?  and if they are, what is the  minimal
set of independent parameters that would have to be adjusted in fitting each
individual source?  In this {\it Letter} we attempt to answer these questions.
Ivezi\'{c} \& Elitzur (1995) showed that IR emission from late-type stars
possesses general scaling properties.  This result was recently extended to
arbitrary systems --- for any type of geometry and density distribution,
infrared emission from radiatively heated dust possesses general scaling
properties when the region's inner boundary is controlled by dust sublimation
(Ivezi\'{c} \& Elitzur 1996; IE96 hereafter).  Thanks to scaling, the number
of free parameters is greatly reduced, enabling a systematic study of the
spectral properties that can and cannot be produced by any family of models.
Here we utilize the scaling approach for detailed modeling of the SEDs
observed in Herbig Ae/Be stars.

\section
                                 {MODELING}

We have developed a numerical code, described in IE96, that solves exactly the
radiative transfer spherical problem taking full advantage of scaling.  Only
two scales need be specified for a complete solution when the shell's inner
radius is controlled by dust sublimation --- the sublimation temperature
\Tsub\ and the overall optical depth at some fiducial, arbitrary wavelength.
For simplicity, we choose \Tsub\ = 1500 K in all sources for both graphite and
silicate grains. The small differences that may exist between these species,
as well as variations of \Tsub\ itself, have a negligible effect on the
results. We choose 0.55~\mic\ as the fiducial wavelength that sets the scale
of optical depth, and the corresponding \tV\ is the only free parameter
allowed to be adjusted in modeling of individual sources.  All other input
properties involve dimensionless, normalized profiles that fall into three
categories:

\paragraph {The external radiation:}   Luminosity is irrelevant.  The
normalized profile $F_\lambda/F$, where $F$ is bolometric flux, is the only
input required for the external radiation. For each source we use the spectral
shape of the Kurucz (1979) model atmosphere corresponding to the spectral type
of the central star.

\paragraph {Dust optical properties:} The magnitude of absorption coefficient
is irrelevant, only the extinction spectral shape $\q = \k/\kappa_V$ and
albedo \al\ enter.  These quantities are determined by the grain chemical
composition and size distribution, and we assume that both properties are
uniform throughout the envelope.  Following Draine \& Lee (1984), the chemical
composition is a mixture of astronomical silicate and graphite grains at the
ratio $n_{\rm Si}:n_{\rm C}$ = 1.12. We take the graphite dielectric
coefficient from Draine \& Lee, the silicate from the more recent study by
Ossenkopf, Henning \& Mathis (1992).  The size distribution is the one
proposed by Mathis, Rumpl \& Nordsieck (1977, MRN hereafter) in which grain
radii obey $a \ga \am$ = 0.005 \mic\ and $a \la \ap$ = 0.25 \mic.

\paragraph {Dust density distribution:} The density scale is irrelevant, as is
the size scale of all geometrical dimensions; both enter as one independent
parameter, the overall optical depth. Only the spatial distribution of dust,
described by a dimensionless, normalized distribution, is required.  Introduce
the dimensionless radial coordinate $y = r/r_1$, where $r_1$ is the dust
sublimation radius.  The shell inner boundary is always at $y = 1$ and the
actual value of $r_1$ never enters.  If the dust density is denoted
$\rho_d(r)$, only the dimensionless, normalized distribution $\eta(y) =
\rho_d(y)/\int_1^\infty \rho_d(y)dy$ is required.  We choose $\eta \propto
y^{-1.5}$, attempting to model all sources with a single density profile.

\subsection                         {Data}

In the recent catalog of Th\' e et al.\ (1994) we have identified all the
Herbig Ae/Be stars with good spectral coverage from 0.3 \mic\ to far-IR
wavelengths, including IRAS data of reliable quality; in some cases we
improved on the listing in the IRAS point source catalog using the ADDSCAN
procedure (Weaver \& Jones 1992). Among these sources we looked for those that
meet the HSVK disk criterion $\lFl \propto \lambda^{-4/3}$ for the underlying
continuum to the longest wavelength observed. We then selected for detailed
modeling those stars which show evidence for the 10 \mic\ silicate feature
commensurate with the standard $n_{\rm Si}:n_{\rm C}$ ratio used here. The
eight sources listed in Table 1, with spectral types from Th\' e et al.
(1994), meet all of these selection criteria and were used in the modeling
presented here. A future detailed paper will present models for additional
objects, including carbon-rich shells.

\placetable{Table1}

Herbig Ae/Be stars show considerable variability. We have carefully studied
all available data pertaining to the light-curves of our sources, including
data from the recent long-term optical monitoring programs of Manfroid et al.\
(1991) and Shevchenko et al.\ (1993). The reddening some Ae/Be stars exhibit
at visual wavelengths during brightness decreases is attributed to occultation
by optically thick dust clumps in their envelopes (e.g.\ Grinin et al.\ 1991).
Two of the sources selected for fitting (KK Oph, WW Vul) display such
variability, and in each case we choose the observations at maximum visual
brightness. Other sources are not known as highly variable stars, and we
simply average their photometric data. In most cases optical and IR
observations are non-simultaneous; for each source we tried to choose data as
contemporaneous as possible.

\section
                              {RESULTS}

Extinction of the stellar radiation is caused both by foreground interstellar
material and the circumstellar shell, and normally it is impossible to
distinguish between the two; correction for interstellar reddening cannot be
performed when the shell effect is not known, and a model cannot be fitted
without reddening-corrected data. Indeed, the procedure used by HSVK to
estimate the interstellar extinction $A_V$ simply ignored the effect of the
circumstellar matter on $B - V$.  However, within the context of a model for
the shell, the two components can be disentangled with the aid of an
additional color excess that is affected also by dust emission. This emission
comes only from the shell\footnote{Interstellar dust emission (cirrus) can
affect only $\lambda \ga$ 60 \mic\ in some sources.} and is significant only
for $\lambda \ga$ 2--3 \mic, since shorter wavelengths require dust hotter
than the sublimation temperature. In figure 1 we show the data points in a
diagram of the color excesses over naked star spectrum for $B - V$ and $V -
[12]$, where [12] is the 12 \mic\ magnitude. Corrections for interstellar
reddening would move data points to the left parallel to the dotted line by
distances proportional to the corresponding $A_V$.  Our models produce a track
in this diagram, drawn with a solid line, originating from the naked star
location (0,0), with distance along the track increasing with \tV. In the
absence of interstellar extinction, data points would lie on this line and
their positions would determine the optical depth of each shell.  The fact
that within errors all the data points do indeed lie on the displayed track
supports our model and indicates that $A_V$ could be negligible for all
members of this sample.  Furthermore, figure 2 shows the spectra calculated
for the optical depths \tVo\ determined for each source from its location,
within the errors, on the track of figure 1.  These models produce
satisfactory fits for the raw data without any correction for interstellar
reddening over the entire spectral range, not just the three wavelengths used
for the determination of \tVo.  This detailed agreement provides further
support for our model.

\placefigure{Figure1}

\placefigure{Figure2}

The large observational error bars and the similar slopes of the reddening
line and our model track for $\EBV \ga 0.1$ introduce ambiguities into the
determination of \tV\ and $A_V$.  For three sources, reddening corrections
performed within the bounds of these ambiguities result in equal or better
fits, also presented in figure 2; in the case of IRAS 16372-2347, this model
noticeably improves the fit around 2--5 \mic. In each case, the spectral
shapes of the two models are identical for $\lambda \ga$ 5 \mic\ because the
shells are optically thin at these wavelengths. Table 1 lists the optical
depths \tVo\ for all sources, and $A_V$ and \tV\ for the three sources whose
shell properties remain slightly ambiguous.

Our models provide some constraints on the dust properties.  The sublimation
temperature can be varied by a few hundred degrees around 1500 K with little
change in the models.  The only minor effect is a slight shift in the
wavelength of the emission peak discernible around 2--3 \mic.  In the case of
KK Oph, this shift improves the fit for $\Tsub \simeq$ 1000 K. The dust
chemistry is primarily constrained by the strength of the 10 \mic\ feature.
For all the sources displayed here, $n_{\rm Si}:n_{\rm C}$ can vary from its
standard value by no more then \about\ 20\% in either direction.  It should be
noted, however, that other sources, which we will analyze in a future detailed
paper, give clear evidence for carbon-rich dust. A primary example is HR 5999,
as noted already by Th\'{e} et al.\ (1996). Grain sizes are primarily
constrained by $U-B-V$ and $K - [12]$ colors.  We find that the distribution
upper limit \ap\ must fall in the range 0.2--1 \mic\ for these sources, and
present in figure 1 the model track when \ap\ is increased to its maximum
allowed value of 1 \mic. The lower limit must obey $\am \la 0.05$.  Both
limits are consistent with the MRN estimates for interstellar grains.

The spectral index of the dust density profile used in all our models is $p$ =
1.5, as expected in free-fall accretion.  Satisfactory fits are produced only
in the rather narrow range $p \simeq$ 1.4--1.6.  At lower values of $p$ the
near-IR bump is overly suppressed, at larger $p$ the far-IR radiation.
However, there are Herbig Ae/Be stars that show strong evidence for other
density profiles, which we will present separately, presumably reflecting
different evolutionary stages.

\section
                                {DISCUSSION}

Our model results, shown in Figure 2, answer the questions posed in the
Introduction:  The SEDs of these sources, primary examples of accretion disks
according to HSVK, are successfully modeled with spherical envelopes that
differ from each other in only one property --- the optical depth \tV, listed
in each panel and in Table 1.  Furthermore, since these SEDs display the 10
\mic\ feature in emission they {\em cannot} be produced by optically thick
accretion disks. Irrespective of geometry, an emission feature is never
produced if the source is optically thick in the corresponding wavelength
regime.  By contrast, the 10 \mic\ optical depths of our spherical models vary
from 0.01 to 0.1.

Because of the inherent scaling properties of the radiative transfer problem,
modeling can determine \tV\ but can never provide information on density scale
or physical dimensions. These quantities can be estimated only if additional
information is provided. For luminosity $L$ in solar units, $r_1 \sim
10^{12}L^{1/2}$ cm (IE96).  The 100 \mic\ radiation originates from regions
where the dust temperature is \about\ 40 K, corresponding to $r \sim
10^4r_1$.  Therefore, a stellar luminosity of order 100 \Lo\ will result in
shell sizes of order $10^4$ AU at 100 \mic, as observed (e.g.\ Di Francesco et
al.).  With standard gas-to-dust mass ratio \rgd\ = 200, the gas column
density outside the sublimation radius $r_1$ is $2\times10^{21}\tV$
cm$^{-2}$.  For free-fall accretion, this column corresponds to $\Mdot =
0.5\times10^{-8}\tV M^{1/2} L^{1/4}$ \Mo\ yr$^{-1}$ where $M$ is the central
mass in solar units.  It should be noted that if grains are depleted in the
envelope, \rgd\ increases and with it the estimate of \Mdot.

Our model envelopes are purely spherical.  By contrast, HKC invoke large axial
cavities in their spherical configurations even though they too attribute the
IR emission to $r^{-3/2}$ density distribution.  HKC present two detailed
model calculations. The one relevant for the sources considered here, because
it displays the 10 \mic\ feature in emission, has \Mdot\ = $10^{-6}$ \Mo\
yr$^{-1}$, corresponding to \tV\ = 30--40; our calculations with such \tV\
reproduce the SED presented by HKC. Because of the large extinction of this
model, the stellar radiation is heavily extinguished by the envelope. To
produce the observed visual fluxes, HKC proposed axial cavities as a means to
obtain a clear line of sight to the star so that in this scheme, the observed
$E(B - V)$ is determined purely by interstellar extinction. The ellipse marked
HKC in Figure 1 shows their model predictions, and all data points should
congregate on the reddening line originating from it.  Instead, all the
sources in our sample fall below this line, more than two standard deviations
away.  If $A_V$ is determined from \EBV, the HKC model predictions exceed the
observed 12 \mic\ flux by more than 2 magnitudes.  One could attempt to modify
the HKC approach with grazing rays, producing a partially obscured star, but
this only exacerbates the problem because then the HKC ellipse moves upward
and to the right in the diagram, away from the data points. The HKC model is
not adequate for the sources in this sample because its optical depth is so
large.  If there is no dust depletion in these shells, the smaller optical
depths required by the data imply mass loss rates 10--100 times smaller than
employed by HKC.

In conclusion, our study amplifies the findings by Di Francesco et al.\ that
extended envelopes, optically thin at IR wavelengths, dominate the IR emission
from group I Herbig Ae/Be stars.  If present, optically thick accretion disks
can only make a negligible contribution to the observed radiation.  Large
axial cavities are not present in the sources modeled here. Similarly,
free-free emission by circumstellar gas, advocated by Berrilli et al., seems
to be superfluous.

\acknowledgments

We are grateful for the permission to use unpublished photometric data by
Drs.\ V.S.\ Shevchenko and W.\ Wenzel. Our model fitting procedure was greatly
aided by a program developed by Charles Danforth. We thank the Center for
Computational Sciences at the University of Kentucky, especially the director
Dr.\ John Connolly, for generous support that made this collaboration
possible. Support by the Exchange of Astronomers Program of the IAU (A.M.),
NSF grant AST-9321847 and NASA grant NAG 5-3010 is gratefully acknowledged.

\newpage

\let\n=\tablenotemark
\let\t=\tablenotetext
\let\ndt=\nodata

\begin{table}[htbp]
\begin{center}
\caption{\hfil Properties of Modeled Stars  \label{Table1}}
\centerline{}
\begin{tabular}{clcrccc}
\tableline \tableline
 IRAS \# &  \hfil Name  & Type  & V   \hfil&  \tVo\phn & $A_V$ &  \tV   \\
        (a)  &  \hfil (b)   &  (c)  & (d) \hfil&   (e) &  (f) &  (g)   \\
\tableline
00403$+$6138 & BD+61\deg154 &  B8    & 10.55  & 2.7 &  0.7  & 1.8   \\
04555$+$2946 &   HD 31648   &  A2    &  7.66  & 0.4 & \ndt  & \ndt  \\
05385$-$0244 &   HD 37806   &  B9    &  7.95  & 0.7 & \ndt  & \ndt  \\
11575$-$7754 &   HD 104237  &  A4    &  6.60  & 0.5 & \ndt  & \ndt  \\
16372$-$2347 &   HD 150193  &  A1    &  8.55  & 1.6 &  1.4  & 0.3   \\
17070$-$2711 &   KK Oph     &  A5    &  9.40  & 1.7 &  0.4  & 1.2   \\
19238$+$2106 &   WW Vul     &  A3    & 10.30  & 1.0 & \ndt  & \ndt  \\
20005$+$0535 &   HD 190073  &  A0    &  7.87  & 0.5 & \ndt  & \ndt  \\
\tableline
\end{tabular}
\end{center}

\tablecomments{(a) IRAS number and (b) other name of modeled stars.  (c)
Spectral type; (d) observed visual magnitude; (e) visual optical depth of the
circumstellar dust shell obtained from the model fit to the data without any
correction for interstellar extinction; (f) interstellar extinction and (g)
optical depth when dereddening the data improves the overall fit.}

\end{table}

\figcaption[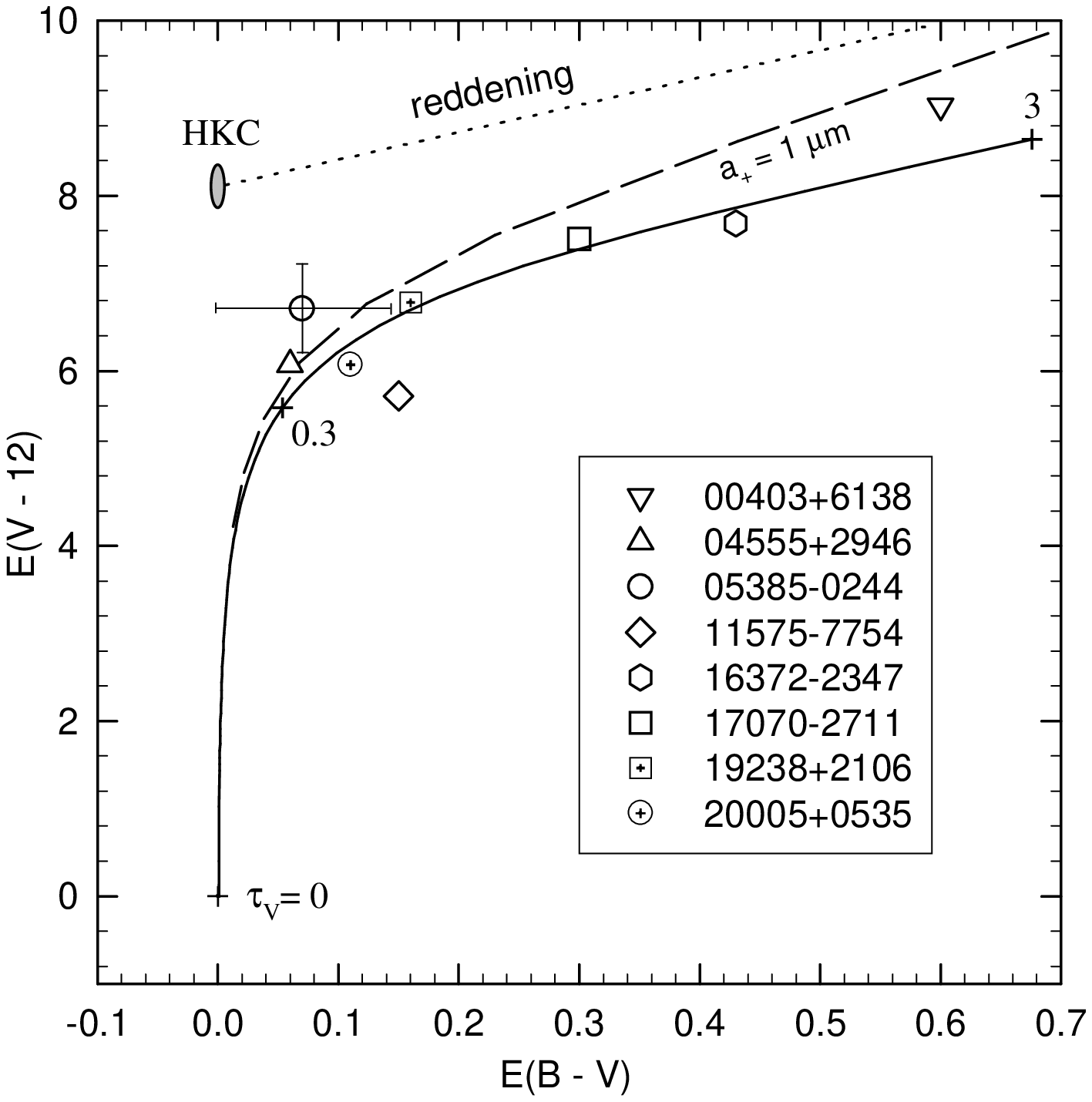]{Color excess diagram for the stars in our sample.
\EBV\ is affected only by the combined effect of interstellar and
circumstellar extinction while \EV\ is affected also by circumstellar dust
emission.  Data points are denoted by the symbols listed in the legend.  Error
bars shown for IRAS 05385-0244 are common to all stars.  Corrections for
interstellar reddening would move each data point to the left parallel to the
dotted line marked ``reddening" by a distance proportional to the
corresponding $A_V$.  The ellipse marked HKC is the prediction of the Hartmann
et al.\ model for cavity opening angles ranging from 30\deg\ to 60\deg. The
solid line is the prediction of our model with MRN dust composition. Position
along this line is determined by the dust shell optical depth \tV, with \tV\ =
0 (a naked star), 0.3 and 3 marked by crosses.  The dashed line is our model
result when the upper limit of the grain size distribution is increased from
0.25 to 1 \mic. \label{Figure1}}

\figcaption[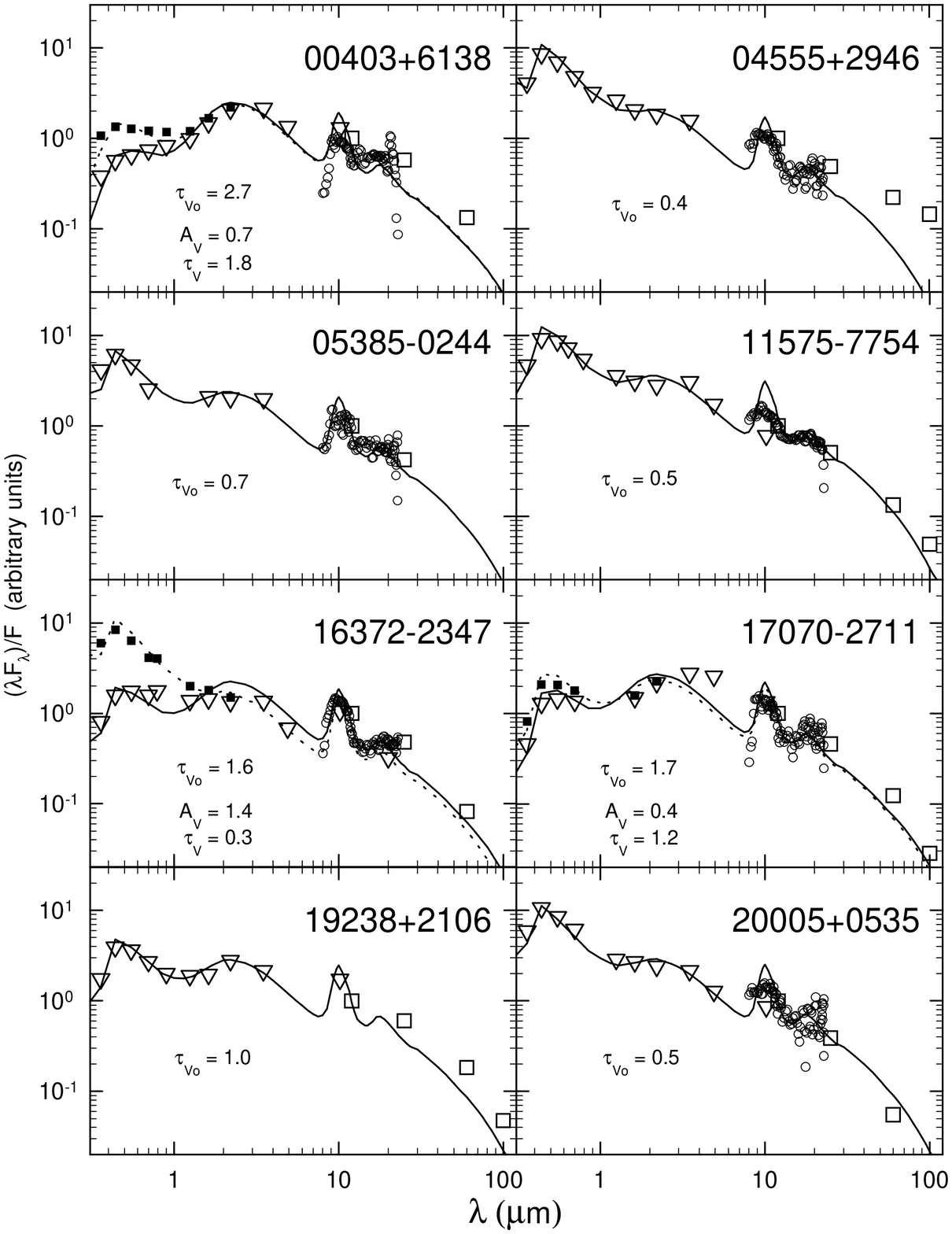]{Observations and model results for the eight stars
listed in Table 1. Each panel is identified by the source IRAS number.  IRAS
data are denoted by circles and open squares for the LRS and PSC listings,
respectively. Triangles denote all other data without any correction for
interstellar extinction. This correction is important only for $\lambda \le$
2.2 \mic, and the full squares in some panels are the same data corrected for
extinction with the listed $A_V$.  The lines are the results of detailed
numerical calculations with the model described in the text. The only envelope
parameter allowed to vary is the optical depth, denoted in each panel. Full
lines correspond to the optical depth \tVo, dotted lines to \tV.
\label{Figure2}}


\newpage

\begin{figure}
\centering \leavevmode \epsfxsize=\hsize \epsfbox[86 172 503 587]{figure1.ps}

\end{figure}

\newpage

\begin{figure}
\centering \leavevmode \epsfxsize=\hsize \epsfbox[20 30 590 765]{figure2.ps}

\end{figure}


\begin{references}

\Ref Berrilli, F., et al.\ 1992, ApJ, 398, 254
\Ref Bertout, C., \& Basri, G. 1991, in The Physics of Star Formation and
      Early Stellar Evolution, ed. C.J. Lada \& N.D. Kylafis (Dordrecht:
      Kluwer), 649
\Ref Di Francesco, J., Evans, N.J., II, Harvey, P.M., Mundy, L.G., \& Butner,
      H.M. 1994, ApJ, 432, 710
\Ref Draine, B., \& Lee, H.M. 1984, ApJ, 285, 89
\Ref Evans, N.J., II, \& Di Francesco, J. 1995, in Disks, Outflows and Star
      Formation, ed. S. Lizano \& J.M. Torrelles, RevMeXAASC, 1, 187
\Ref Grinin, V.P., Kiselev, N.N., Chernova, G.P., Minikulov, N.Kh., \&
      Voshchinnikov, N.V. 1991, Ap\&SS 186, 283
\Ref Hanner M.S., 1988, Infrared Observations of Comets Halley and
      Wilson and Properties of the Grains (NASA89-13330), 22
\Ref Hartmann, L., Kenyon, S.J., \& Calvet, N., 1993, ApJ, 407, 219 (HKC)
\Ref Harvey, P.M., Lester, D.F., Brock, D., Joy, M., 1991, ApJ, 368, 558
\Ref Hillenbrand, L.A., Strom, S.E., Vrba, F.J., \& Keene, J. 1992, ApJ, 397,
      613 (HSVK)
\Ref Ivezi\' c, \v Z., \& Elitzur, M., 1995, ApJ, 445, 415
\Ref Ivezi\' c, \v Z., \& Elitzur, M., 1996, MNRAS, submitted (IE96)
\Ref Kurucz, R.L., 1979, ApJS, 40, 1
\Ref Manfroid, J., Sterken, C.,  Bruch A., et al. 1991. ESO Scientific Report,
      No. 8
\Ref Mathis, J.S., Rumpl, W., \& Nordsieck, K.H., 1977, ApJ, 217, 425 (MRN)
\Ref Ossenkopf, V., Henning, Th., \& Mathis, J.S. 1992, A\&A, 261, 567
\Ref Shevchenko, V.S., Grankin, K.N., Ibragimov, M.A., Melnikov, S.Yu., \&
      Yakubov, S.D. 1993. Ap\&SS, 202, 121
\Ref Shu, F., Adams, F.C., \& Lizano, S. 1987, ARA\&A, 25,23
\Ref Th\' e, P.S., P\' erez, M.R., \& de Winter, D. 1994, A\&AS, 104, 315
\Ref Th\' e, P.S., P\' erez, M.R., Voshchinnikov, N.V., \&
      van den Ancker, M.E. 1996, A\&A, in press
\Ref Weaver, Wm.B., \& Jones, G. 1992, ApJS, 78, 239



\end{references}
\end{document}